\documentclass[usenatbib]{mnras}
\usepackage{graphicx}
\graphicspath{{./../../figures/}}
\usepackage{float}
\usepackage{amssymb}
\usepackage{color}
\usepackage{multirow}
\usepackage[T1]{fontenc}

\newcommand{\sarc}{$^{\prime\prime}\!\!$}
\newcommand{\myreferences}{mybib.bib}

\newcommand{\az}{$\alpha\,$-$\,z$}

\title[Investigating the \az\ relation]{Investigating the cause of the \az\ relation}
\author[Leah K. Morabito]{\parbox{\textwidth}{Leah K. Morabito$^{1}$\thanks{E-mail: morabito@physics.ox.ac.uk} and Jeremy J. Harwood$^{2}$
\\}\\
$^{1}$Astrophysics, University of Oxford, Denys Wilkinson Building, Keble Road, Oxford, OX1 3RH \\
$^{2}$Centre for Astrophysics Research, School of Physics, Astronomy and Mathematics, University of Hertfordshire, College Lane,\\Hatfield, AL10 9AB, UK \\}

\definecolor{Mygrey}{gray}{0.75}

\begin{document}
\date{}
\pagerange{\pageref{firstpage}--\pageref{lastpage}} \pubyear{2014}
\maketitle

\label{firstpage}

\begin{abstract}
The correlation between radio spectral index and redshift has long been used to identify high redshift radio galaxies, but its cause is unknown. Traditional explanations invoke either $(i)$ intrinsic relations between spectral index and power, $(ii)$ environmental differences at high redshift, or $(iii)$ higher inverse Compton losses due to the increased photon energy density of the cosmic microwave background. In this paper we investigate whether the increased inverse Compton losses alone can cause the observed spectral index -- redshift correlation by using spectral modelling of nearby radio galaxies to simulate high redshift equivalents. We then apply selection effects and directly compare the simulated radio galaxy sample with an observed sample with sufficient redshift coverage. We find excellent agreement between the two, implying that inverse Compton losses and selection effects alone can largely reproduce the observed spectral index -- redshift correlation.
\end{abstract}


\section{Introduction}
High redshift radio galaxies (HzRGs; $z\geq 2$) are unique laboratories for studying the formation and early evolution of the massive galaxies, rich clusters and massive black holes we observe in the local Universe. They have extended jets on kpc scales that emit synchrotron radiation detectable in the radio regime. The host galaxies have clumpy optical morphologies \citep{pentericci_vla_2000} and optical spectra indicative of extreme star formation and large stellar masses. They are often found in protocluster environments  \citep{pentericci_vla_2000} and are thought to evolve into present-day dominant cluster galaxies \citep{best_nature_1997,miley_distant_2008}. Less than 200 HzRGs are presently known \citep{miley_distant_2008}, and the highest redshift radio galaxy to date is at $z=5.19$ \citep{van_breugel_radio_1999}. 

Almost all of these HzRGs were found by searches for ultra steep spectrum (USS; defined as $\alpha \lesssim -1.2$ where flux density is $S \propto \nu^{\alpha}$) sources in radio surveys. \citet{tielens_westerbork_1979} first recognized that USS sources were three times less likely to have an optically identified host galaxy, and that their smaller angular sizes implied they were at larger distances. \citet{blumenthal_spectral_1979} found that spectral index did indeed correlate with redshift, with steeper spectral indices associated with objects at higher redshift. Since then, searching for USS sources has been an effective way to identify candidate high redshift sources \citep[e.g.,][]{rottgering_samples_1994} that can be followed up with spectroscopic confirmation. 

While the \az\ correlation is useful for identifying high redshift galaxies, it is not understood what causes the relation. The traditional explanation is that the observed USS is due to a radio $k$-correction coupled with the fact that fixed observing frequencies probe higher rest frame frequencies of the radio spectra for higher redshift sources, where steepening due to synchrotron losses is more pronounced. The high-frequency spectra will also be impacted by losses due to inverse Compton scattering of cosmic microwave background (CMB) photons \citep{krolik_steep_1991}. However, \citet{klamer_search_2006} investigated the rest-frame radio spectra of 37 USS HzRGs with matched-resolution observations spanning $2.3-6.2\,$GHz and found that the $k$-correction did not impact the overall relation between spectral index and redshift. 

Two alternative explanations have been proposed. The first is that higher ambient density could cause steeper electron energy spectra in the particle acceleration processes at the jet working surfaces. Higher ambient density is expected at higher redshifts, and the radio spectra of HzRGs could therefore be steeper than local radio galaxies \citep{athreya_redshift_1998,klamer_search_2006}. The attraction of this explanation is that it could result in both an \az\ relation and an $\alpha-$luminosity relation. However, recent work by \citet[][and references therein]{vernstrom_radio_2017}  using rotation measures show that there is no statistical difference in the environments of HzRGs and their lower-redshift counterparts. 
 
Another explanation is that the \az\ relation arises naturally from a correlation between $\alpha$ and luminosity \citep{chambers_4c_1990,blundell_inevitable_1999}. For models where higher jet powers produce steeper integrated spectra, when taking Malmquist bias into account no intrinsic relation between \az\ is necessary to match observations. It is difficult to study these kinds of effects observationally with flux density limited surveys. \citet{ker_new_2012} examined the relevant relationships amongst power, linear size, redshift, and spectral index for both low and high frequency selected surveys separately, finding only a weak \az\ relation which is dominated by the scatter in $\alpha$. Their findings are consistent with increasing ambient density at higher redshifts driving the \az\ relation.  

In this paper we focus on understanding whether the redshift-dependent inverse Compton losses alone are enough to reproduce the observed \az\ relation. This is a new approach which makes use of the Broadband Radio Astronomy ToolS software package \citep[\textsc{brats}\footnote{http://www.askanastronomer.co.uk/brats};][]{harwood_spectral_2013,harwood_spectral_2015} to extract physical properties of low-redshift radio galaxies. Assuming that HzRGs are intrinsically the same as their local counterparts \citep[supported by the findings of][]{morabito_lofar_2016}, we then simulate a population of radio galaxies up to $z\sim6$ letting only the inverse Compton losses vary. We then apply selection effects and compare with an observed sample to determine if the observed \az\ relation can be reproduced in this way alone. 

In Section~\ref{sec:s2} we describe the initial sample. In Section~\ref{sec:s3} we explain how we use \textsc{brats} to model the data, and introduce selection effects. We describe the construction of our simulated high-$z$ sample in Section~\ref{sec:s4} Results are presented in Section~\ref{sec:s5} followed by discussion and conclusions in Section~\ref{sec:s6}. Throughout the paper we assume a $\Lambda$CDM concordance cosmology with $H_0=67.8$ km$\,$s$^{-1}\,$Mpc$^{-1}$, $\Omega_{\textrm{m}}=0.308$, and $\Omega_{\Lambda}=0.692$, consistent with \citet{planck_collaboration_planck_2016}. Spectral index is defined as $\alpha$ with flux density $S\propto\nu^{\alpha}$. 

\section{Initial Sample}
\label{sec:s2}
Integrated flux density measurements are required across a wide range of frequencies to perform robust spectral modelling  by fitting any curvature present in the overall SED. Extremely low frequencies ($\lesssim 100\,$MHz) can be impacted by free-free absorption and/or synchrotron self-Compton absorption, and will not be well-fitted by spectral aging models that do not include these physical processes. At frequencies $\gtrsim10\,$GHz, most sources will be highly resolved and integrated flux density measurements will be unreliable. The frequency range of interest is therefore $\sim 100\,$MHz to $10\,$GHz. 

The NASA/IPAC Extragalactic Database\footnote{The NASA/IPAC Extragalactic Database (NED) is operated by the Jet Propulsion Laboratory, California Institute of Technology, under contract with the National Aeronautics and Space Administration.}  provides easy access to collated information from many different surveys. We searched NED for sources with the following constraints:
\begin{itemize}
\item Classified as galaxy, quasar, or unclassified radio source. 
\item Spectroscopic redshift available and $z<1$.
\item Radio photometry available at 178$\,$MHz.
\end{itemize}
The final criteria ensures that the source has data from the 3CR survey \citep{laing_bright_1983}, a complete spectroscopic survey containing the type of bright, jetted radio sources in which we are interested. We added data points from the new TIFR GMRT Sky Survey Alternative Data Release 1 \citep[TGSS ADR1; an all-sky survey at 150$\,$MHz;][]{intema_gmrt_2017}, by cross-matching radio sources within half of the average beam size of the survey (25\sarc\ ). We also added the complete catalogue from the Westerbork Northern Sky Survey \citep[WENSS;][]{rengelink_westerbork_1997} at 325$\,$MHz since NED only has information from preliminary catalogues. We used a cross-matching radius of 27\sarc\ . At this point we discarded sources with fewer than five measurements between 70$\,$MHz and 10$\,$GHz to ensure that that spectrum is well constrained, yielding an initial sample of 770 sources. 

Some data contained in NED is inappropriate for this study, e.g. measurements from spectral line studies and resolved radio galaxy components. We used keyword searches to automatically remove individual measurements which were indicated to be other than integrated flux density\footnote{Keywords: `peak', `core', `lobe', `line', `vlbi', `very long baseline', `MERLIN'.}. A later visual inspection identified a further 15 surveys with measurements other than integrated flux density, and we removed these data as well. We completely removed sources from the sample which were marked as either flat-spectrum or blazars\footnote{Keywords `flat-spectrum', `blazar', `bl lac'.}, which can have strong orientation effects that would bias the spectral modelling. There were 10 surveys found from visual inspection which targeted compact steep spectrum, Gigahertz-peaked spectrum, and flat-spectrum sources; we also completely removed these sources from our sample. Eleven more sources with visually identified low-frequency turnovers were also removed.

Additionally we removed all data from three surveys/instruments, the Precision Array for Probing the Epoch of Reionization \citep[PAPER;][]{jacobs_new_2011}, Radio sources observed with the Culgoora array \citep{slee_radio_1995}, and  the Texas Survey of radio sources \citep{douglas_texas_1996}. Visual inspection showed these measurements to be consistent outliers (from data points within $\Delta \nu/\nu \lesssim$0.15). The resolution of all these instruments is at least a few arcminutes, and the measurements may be either probing larger scales of radio emission or confused with nearby sources. This is consistent with the outliers in the case of PAPER and Culgoora measurements, which were always at higher flux densities than the surrounding data points. The Texas outliers mostly tend towards lower flux densities than surrounding data points, but it is known that the flux density measurements are impacted by significant systematic error \citep[for more details see][]{douglas_texas_1996}. 

At this point we more carefully checked the number of spectral points per object.  A spread of at least five measurements at distinctly different frequencies is required for SED fitting. In practice, this generally means a minimum of two flux density measurements at frequencies $\lesssim500\,$MHz and three measurements $\gtrsim500\,$MHz, where curvature will be most prominent. Radio observations are limited to the discrete frequency bands of receivers on the telescopes, and it is common for multiple observations to exist at the same observing frequency. We therefore imposed the following requirements:
\begin{itemize}
\item There are at least two measurements $<500\,$MHz which have $\Delta \nu \geq100\,$MHz .
\item A simple power-law fit to data $<500\,$MHz must be steeper than $-0.5$ to avoid low-frequency turnovers.
\item There are at least three measurements above 500$\,$MHz.
\item If all three measurements $>500\,$MHz are $<2\,$GHz, the largest $\Delta \nu$ is $\geq 500\,$MHz.
\item If the maximum observing frequency is $>2\,$GHz, the largest $\Delta \nu$ for data above $>500\,$MHz must be $\geq 1\,$GHz.
\end{itemize}
The final sample included 448 sources. 

\section{Modelling the data}
\label{sec:s3}

The most commonly used model for determining the age of a source using only the integrated flux density is the Continuous Injection (CI) model first proposed by \cite{jaffe_dynamical_1973}. This model is based on the synchrotron and inverse Compton losses described by the single injection Jaffe and Perola model which gives the expected emission for a single age region of plasma (see \citealt{harwood_spectral_2013,harwood_spectral_2015} for a detailed description of these models, and \citealt{longair_high_2011} for a full derivation). To allow for the integrated flux of a source to be considered, the CI model in addition assumes an injection of fresh, zero-age plasma and integrates over all ages for the lifetime of the source.

It is worth noting that the CI model is unable to provide a robust measure of the intrinsic age of radio galaxies \citep{harwood_spectral_2017}, especially since various age distributions of electrons in the lobes are mixed in the observations. \cite{turner_raise_2018} find that the CI model describes a majority of powerful radio AGN over a broad frequency range, supporting our argument that the CI model does still provide a good description of the general shape of the spectrum when sparsely sampled in frequency space providing a characteristic, if not intrinsic, source age. As here we are only interested in the effect of inverse-Compton losses, rather than the age of the population, this is unlikely to significantly impact our results.

The CI model assumes continuous injection of fresh particles, and therefore we are biased against remnant source population, where there is no longer ongoing injection. \cite{mahatma_remnant_2018} estimate an upper limit of 9 percent to the remnant population, which could contaminate the final observed sample with which we compare our simulations. However, remnant sources have low surface brightness and will not be detected at high redshift. There are over 100 sources at $z < 0.5$ in the observed sample, and the results will be dominated by the non-remnant population. 

To determine the initial electron energy distribution (injection index) of the sources we used the \textsc{brats} ``findinject'' function \citep{harwood_spectral_2013} with a search range of $\alpha_{inj} = -0.5$ to $-1.0$ in steps of $-0.01$, and assuming a fixed magnetic field strength of $1\,$nT. This method allows for any curvature within the observed spectrum to be accounted for and so if not dependent on being determined well below the spetcral break, providing more robust values that those determined from a simple power law taken at the lowest available frequencies. While there is almost certainly a range of magnetic field strengths within our sample, the radiative losses (hence observed spectral index) are proportional to both the magnetic field strength of the source and the equivalent field of the CMB where $B = \sqrt{B_{lobe}^{2} + B_{CMB}^{2}}$. Due to the equivalent field strength scaling as a function of redshift such that $B_{CMB} = 0.318 (1 + z)^{2}\,$nT, it is this term which becomes increasingly significant at high redshifts. It is these sources which most strongly influence the \az\ relation and so the effect of any variations in our initial sample should not impact significantly on our overall result.

\begin{figure*}
\includegraphics[width=\textwidth,clip,trim=0cm 0cm 1cm 0cm]{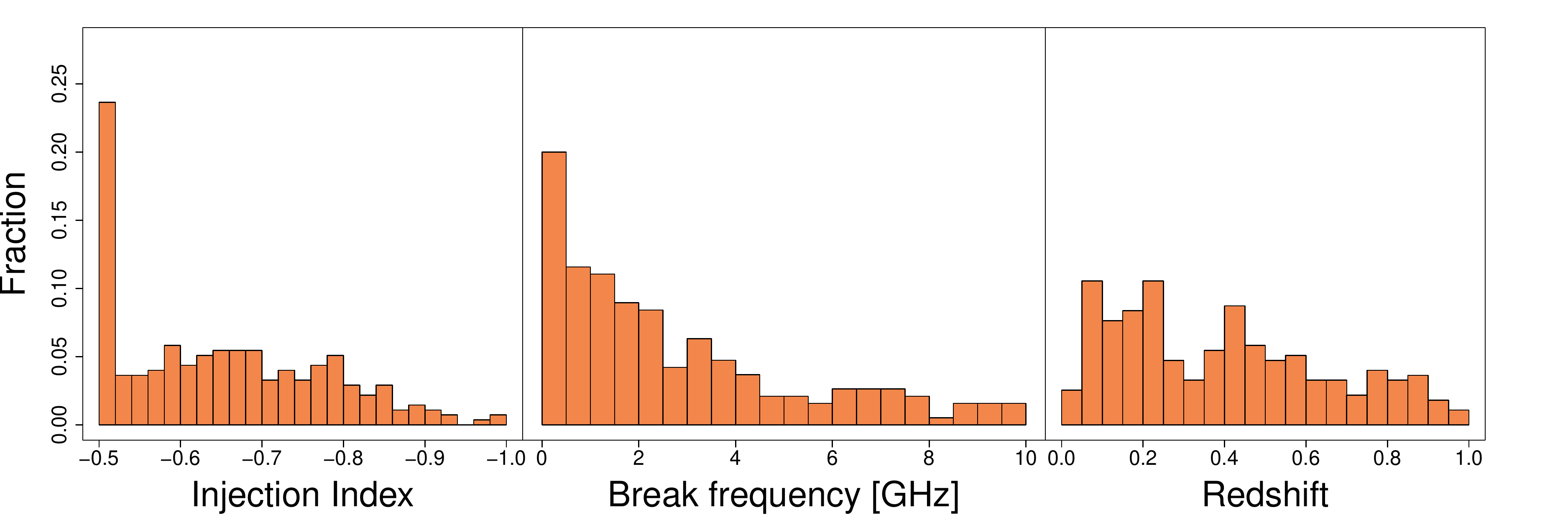}
\caption[Distributions of initial sample parameters.]{\label{fig:f1}Distributions of initial sample parameters. From left to right: injection index, break frequency, and redshift. The distribution of spectral ages is not plotted, but is related to the break frequency as $t\propto \nu^{-0.5}$.}
\end{figure*}

The results of this spectral modelling includes the spectral age, CMB magnetic field energy density, break frequency in the spectrum, injection index, $\chi^2$ of the fit, and confidence level of the fit. We determined the best-fit injection index for each source by taking only fits with confidence levels greater than 1$\sigma$, and finding the best $\chi^2$ value. We also removed 26 sources for which the spectral modelling produced ages consistent with zero. This left a total of 288 sources in the initial sample, and we show the distributions of parameters associated with the fits in Fig.~\ref{fig:f1}. There is a noticeable peak of injection index values around $-0.5$, the upper limit of the injection fitting, which is the commonly assumed physical limit for first-order Fermi acceleration. While many of these sources will be FR-Is where $\alpha_{inj} \approx -0.5$,  and we have made efforts to remove sources where a turn over in the spectrum due to e.g. self-absorption is present, contamination in this bin could still exist from sources that have begun to turn over, flattening below this physical limit. In any case, as sources where the model is poorly fitted are excluded from our sample this effect should be minimal and such contamination is not likely to impact greatly the analysis in the work.

\begin{figure}
\includegraphics[width=\columnwidth]{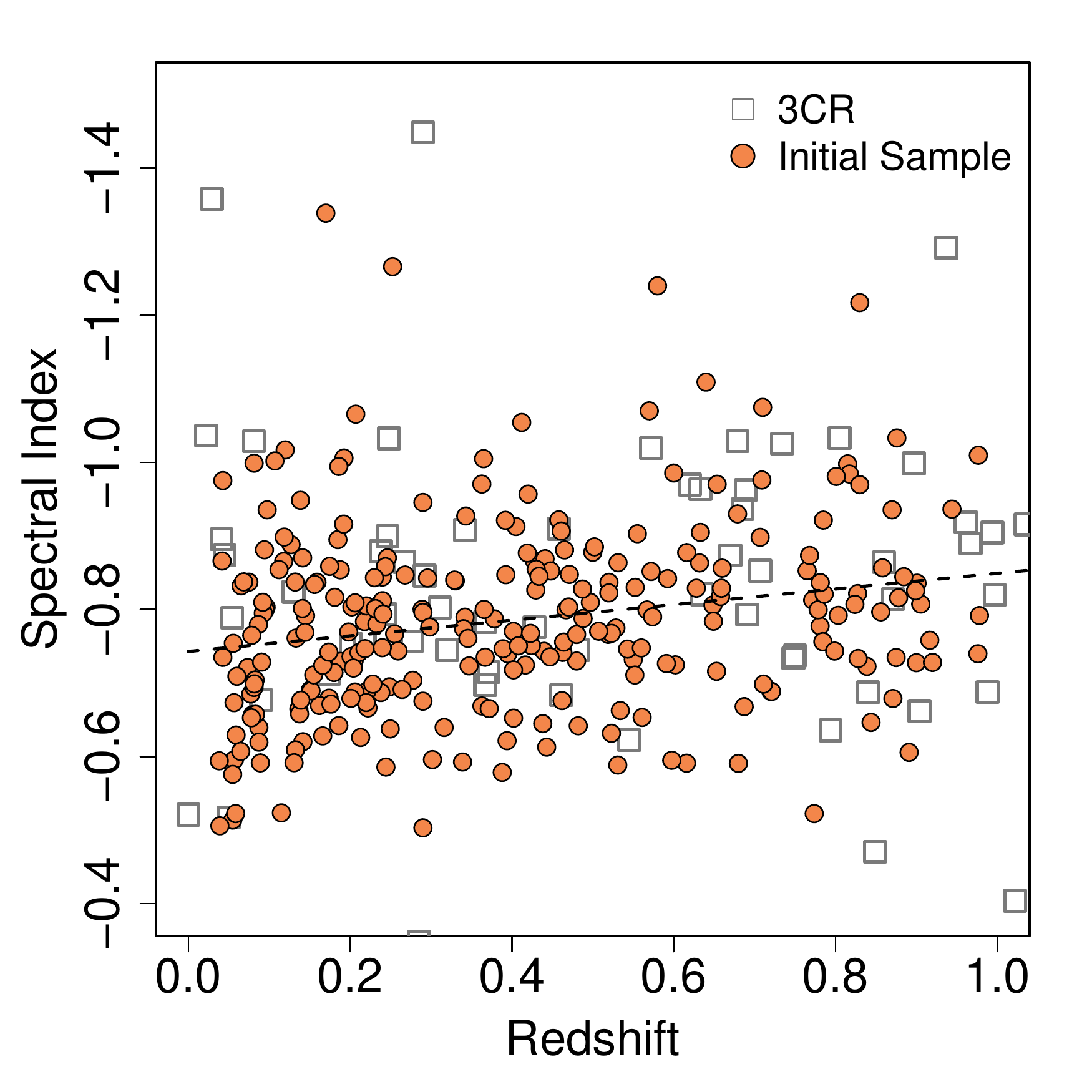}
\caption[\az\ for the initial sample]{\label{fig:f2}The \az\ relation for the initial sample (orange circles), which is limited to $0.5\leq z \leq 1$. A linear fit to the data (black dashed line) shows a slope of $0.140\pm0.029$. We also plot the \az\ relation for the 3CR sample as gray points.}
\end{figure}

\textsc{brats} also outputs rest frame model spectra, that were used to measure the \az\ relation for our initial sample. \citet[][hereafter DB00]{de_breuck_sample_2000} calculated $\alpha$ using fixed observing frequencies of 325 or 365$\,$MHz for the low-frequency point, and 1.4$\,$GHz for the high frequency point. Here we use fixed observing frequencies of 325$\,$MHz and 1.4$\,$GHz to calculate $\alpha$. Fig.~\ref{fig:f2} shows the \az\ relation for the initial sample, using the rest-frame model spectra. For comparison, we also show the 3CR sample from \cite{spinrad_third_1985}, matched with WENSS to provide the spectral index between 325$\,$MHz and 1.4$\,$GHz. This sample is therefore slightly different than that used in our model fitting, but it is instructive to show that our initial sample spans the same type of parameter space. A linear fit to the initial sample shows a slope $0.140\pm0.029$. The slight evolution with redshift is likely due to an evolution of spectral index with radio power \citep{chambers_4c_1990,blundell_inevitable_1999}, coupled with the fact that the data was drawn from flux-limited surveys. This will be addressed later on by imposing flux cuts on the final simulated sample. 

\section{Constructing a High-redshift Sample}
\label{sec:s4}
We constructed a high-redshift sample by matching the observed distribution of redshifts from DB00. We fit a power-law to the observed redshift distribution and use the acceptance-rejection method to simulate $10^6$ redshifts consistent with this distribution (Fig.~\ref{fig:f3}). Drawing redshifts from this simulated distribution to match the observed number of sources and redshift distribution in the DB00 sample, we randomly assign sources to each simulated redshift. This means some of the sources from the initial sample are used more than once, but for different simulated redshifts. To avoid being biased from a single random simulation, we performed 500 Monte-Carlo simulations. 

\subsection{\textit{k}-correction}
\begin{figure*}
\includegraphics[width=\columnwidth,clip,trim=0cm 0cm 0cm 0cm]{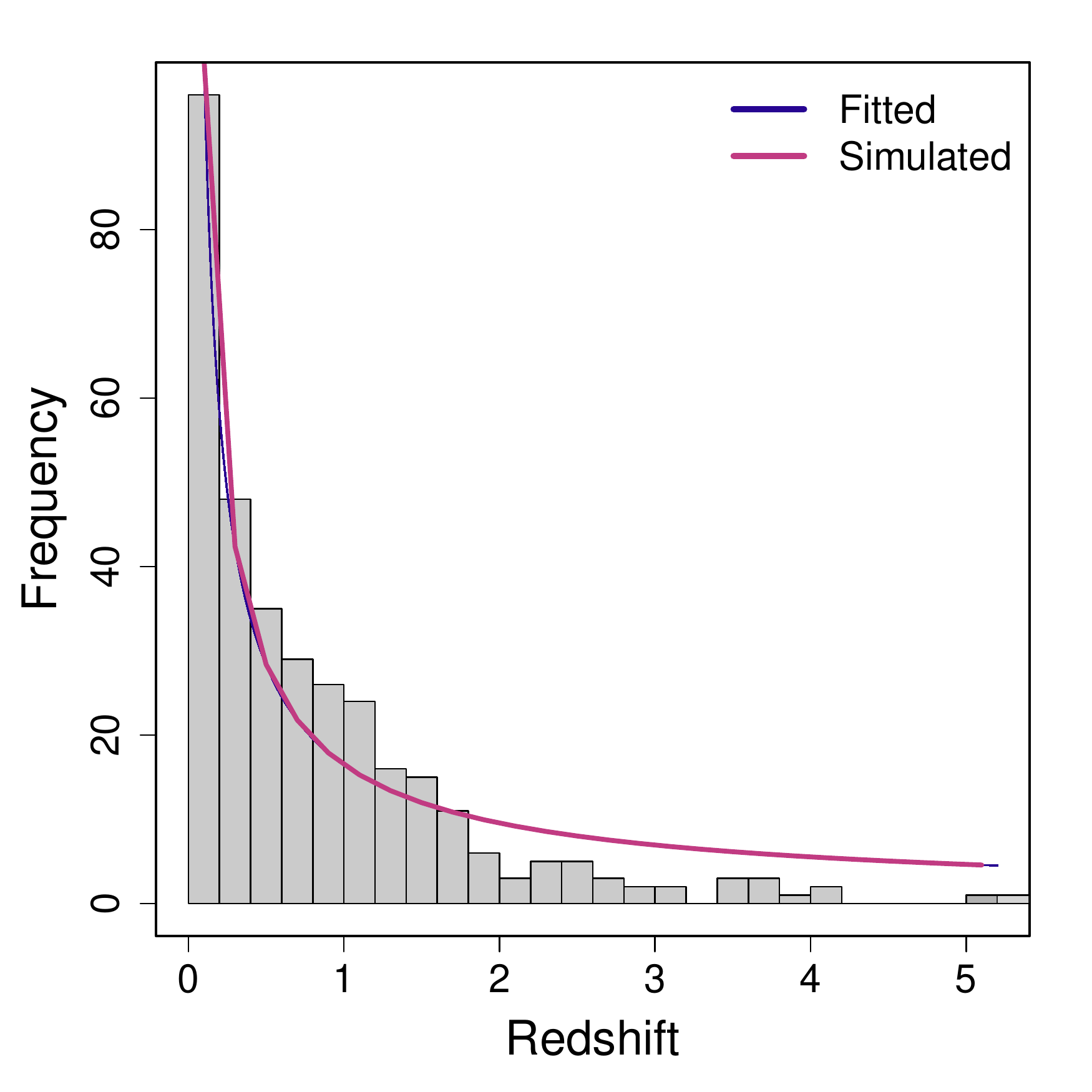}
\includegraphics[width=\columnwidth,clip,trim=0cm 0cm 0cm 0cm]{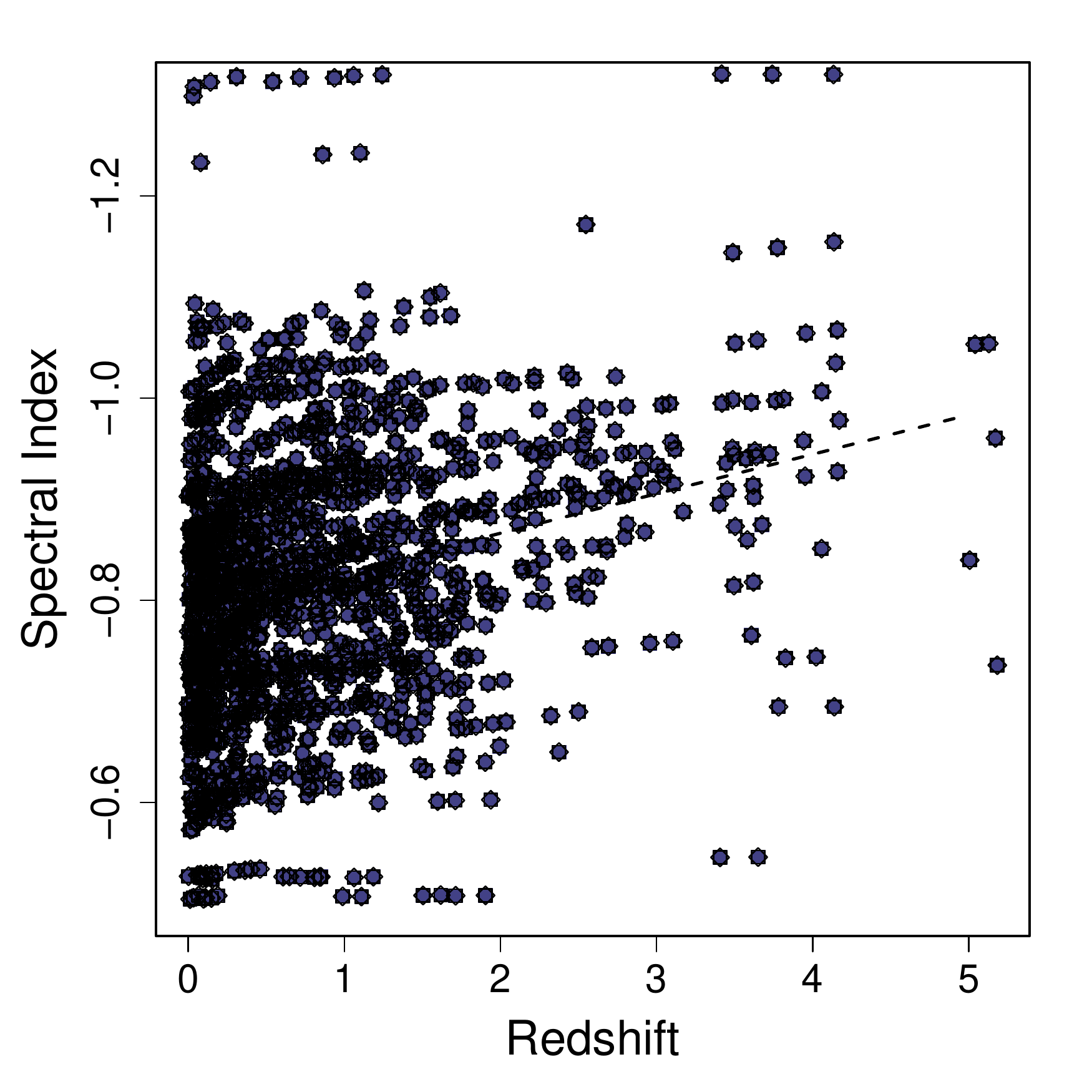}
\caption[$k$-corrected sample and distributions of redshifts]{\label{fig:f3} \emph{Left:} The redshift distribution from DB00. The blue line is a fit to the redshift distribution, and the red line shows our simulated redshifts. The redshifts for the simulated sample are drawn from the simulated distribution, taking care to match the redshift distribution in the DB00 sample. \emph{Right:} The sample of local galaxies that have been $k$-corrected with their simulated higher redshifts, drawn from the redshift distribution on the left. }
\end{figure*}

We first modelled the effects of the $k$-correction. For this, we took the high-redshift sample and $k$-corrected the rest-frame spectra to the observed frame using the simulated redshifts. We then measured the observed model spectra at fixed frequencies of 325$\,$MHz and 1.4$\,$GHz, and performed a linear fit to measure the slope of the relation between $\alpha$ and $z$. The redshift distribution and observed \az\ relation are shown in Fig.~\ref{fig:f3}. The linear fit to the data shows the slope of the relation is now $0.041\pm0.010$, which is slightly flatter than the initial sample. 

\subsection{Inverse Compton from CMB}
To explore the effects of inverse Compton scattering of the CMB, the \textsc{brats} model extrapolation algorithm \citep[described in][]{harwood_spectral_2013} was used to simulate a radio galaxy spectrum for each source in the simulated redshift distribution. Fixing the ages and injection index to those determined by the initial model fitting ensures that the only difference in the resultant spectra are the effects of inverse Compton scattering. We again $k$-corrected the rest-frame model spectra to the observed frame and measured the spectral index between 325$\,$MHz and 1.4$\,$GHz. 

\subsection{Observational Biases}
To compare with the sample from DB00 we must be careful to use only objects from the simulated sample that would satisfy the selection effects in the observed sample. The 3CR and Molonglo Reference Catalogue (MRC) samples are complete down to their flux density limits, and comprise much of DB00 (although DB00 use other samples of USS sources selected from larger, incomplete samples). We therefore select objects to be included in the `observed' simulated sample if they satisfy either of the following criteria:
\begin{itemize}
\item The source is above the flux density limit of the 3CR sample (10$\,$Jy) with a redshift $\leq z_{max}$ of the 3CR sample ($z_{max}=2.474$).  
\item The source is above the flux density limit of the 4C sample (2$\,$Jy) with a spectral index at least as steep as $\alpha=-1.03$, the limiting spectral index for this sample in DB00.
\end{itemize}

\section{Results and Discussion}
\label{sec:s5}
\begin{figure*}
\begin{minipage}{0.7\textwidth}
\begin{center}
\textbf{\Large Observed + Simulated} \\[-7pt]
\includegraphics[width=\textwidth,clip,trim=0cm 0cm 0cm 0cm]{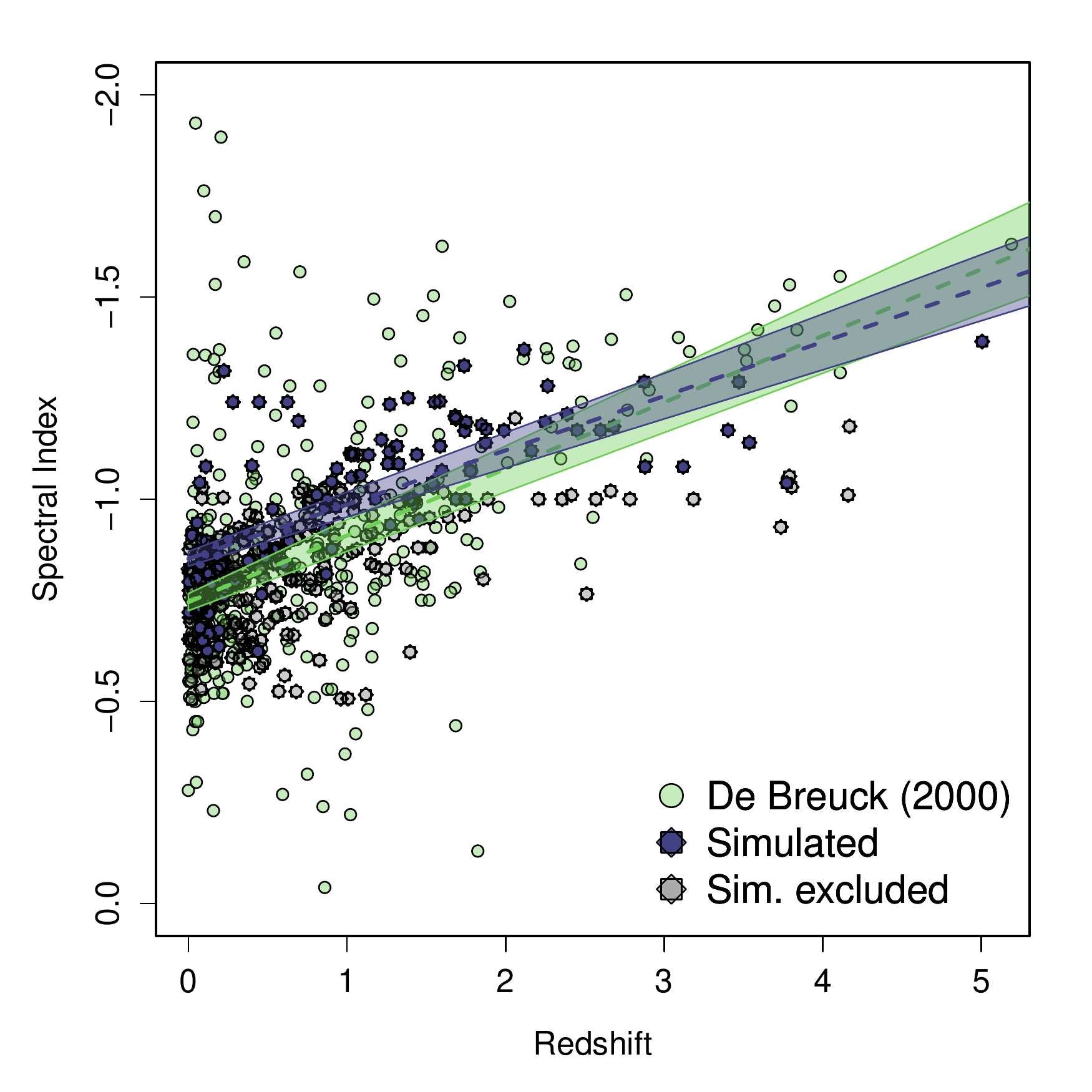}
\end{center}
\end{minipage}%
\begin{minipage}{0.35\textwidth}
\begin{center}
\textbf{\Large Observed} \\
\includegraphics[width=0.9\textwidth,clip,trim=0cm 0cm 0cm 0cm]{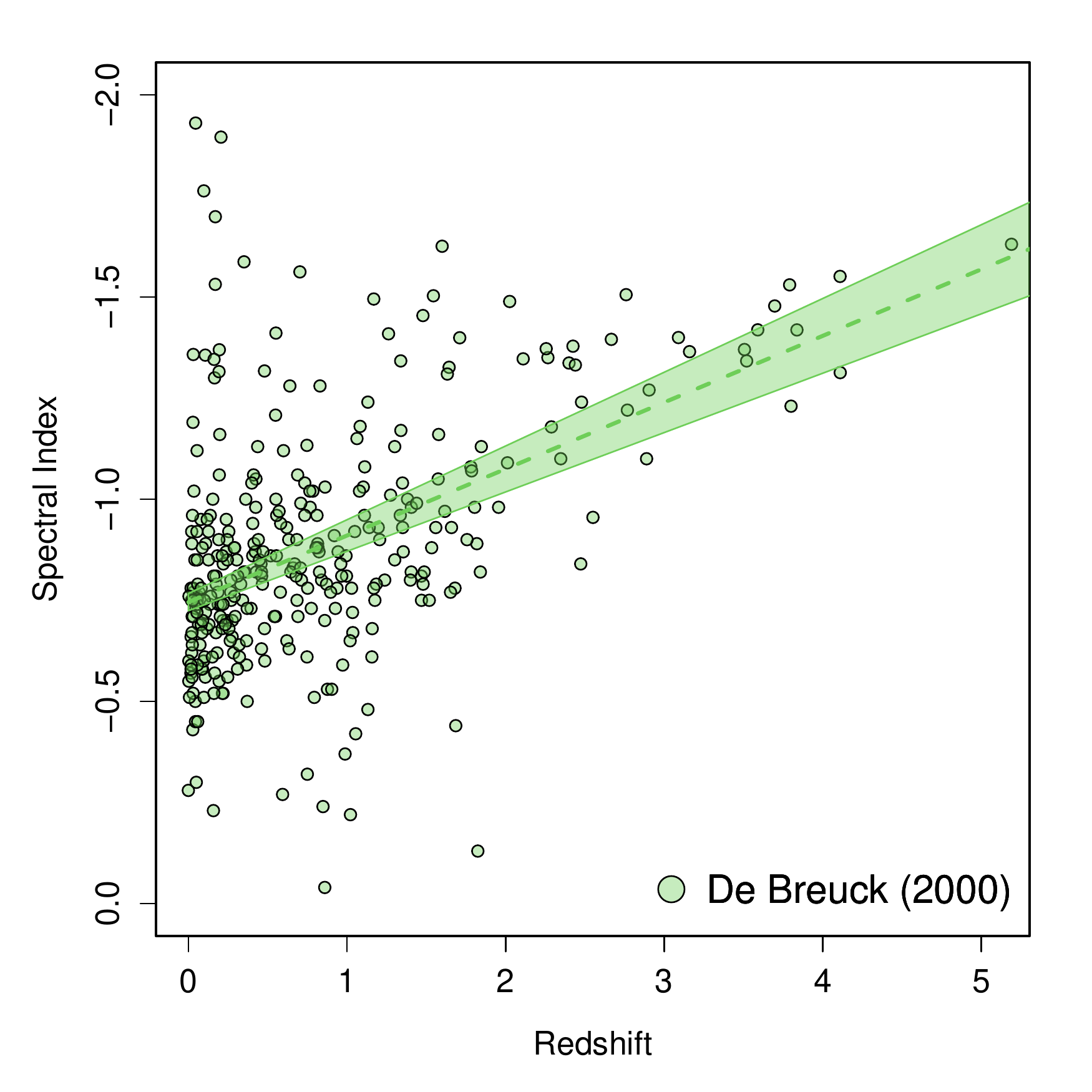}\\
\textbf{\Large Simulated} \\
\includegraphics[width=0.9\textwidth,clip,trim=0cm 0cm 0cm 0cm]{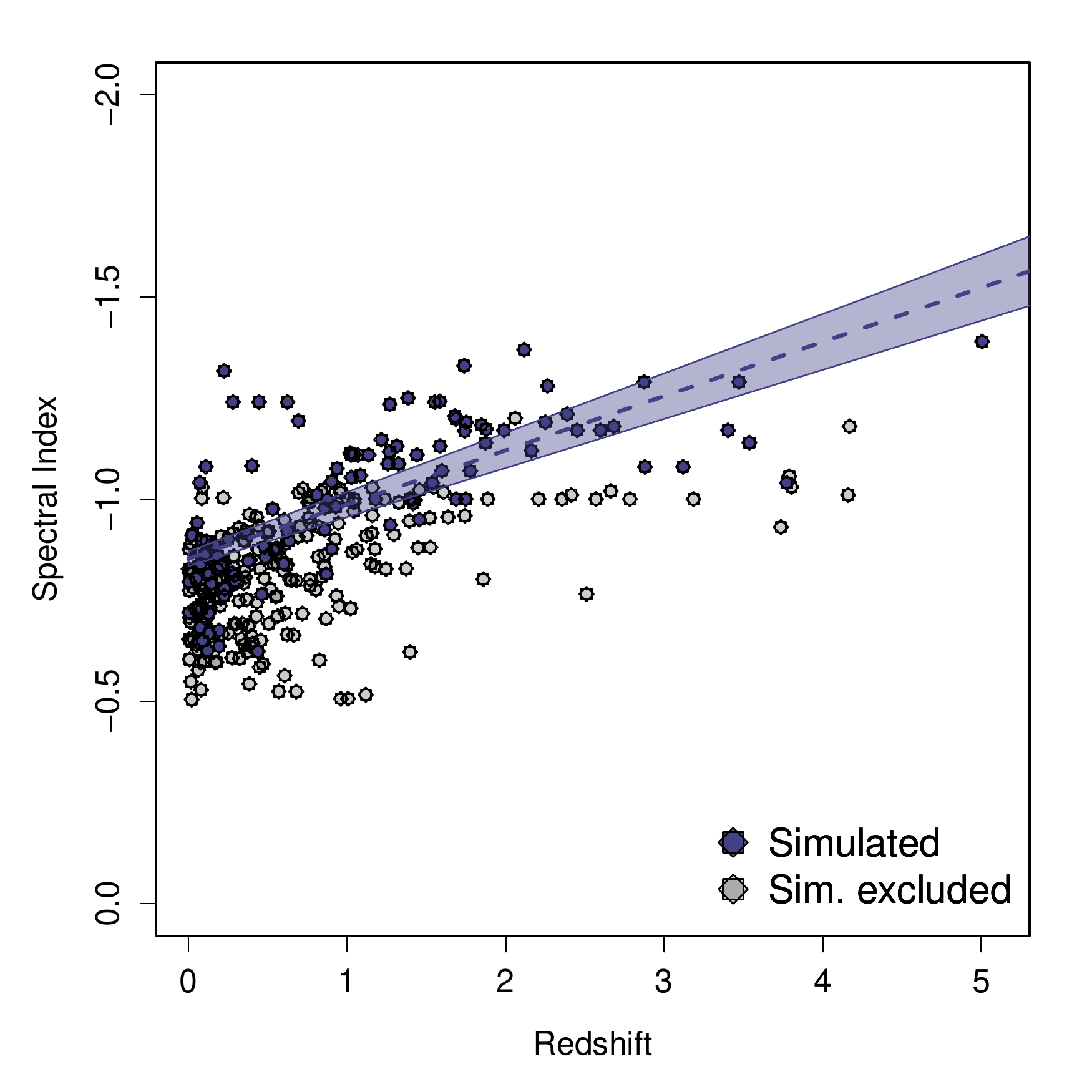}
\end{center}
\end{minipage}
\caption[The \az\ for the observed and simulated samples]{\label{fig:f4}The \az\ for the observed and simulated samples. The large panel on the left shows both observed and simulated samples together, while the panels on the right show the observed (\emph{top}) and simulated (\emph{bottom}) samples separately. Light green circles represent the \cite{de_breuck_sample_2000} and include data points from the 3CR survey, the 4C survey, and new additions from that study. A linear fit to the observed sample is shown by the light green dashed line, with the light green shaded area representing the fit errors. The simulated data which meet the selection criteria explained in Section~\ref{sec:s3} are plotted as purple stars, with a linear fit shown by a purple line with the corresponding uncertainties in the shaded red area. Light gray stars represent the simulated data that did not meet the selection criteria and were excluded from the linear fit.}
\end{figure*}

We plot the final results in Fig.~\ref{fig:f4} for the best-fit Monte-Carlo simulation. Shown in the figure are the observed sample from DB00, the simulated sample with inverse Compton effects modelled in the spectra, and the sample which also matches the selection criteria described in the last section. The linear fits of the simulated sample matching the selection criteria and the observed sample are also shown. The distributions of the fitted slope and intercept from the Monte-Carlo simulations were Gaussian, and we take the mean and standard deviation as our final values and uncertainties. We find $\alpha = -(0.13\pm0.014)z -0.86\pm0.02$ for the simulated sample including selection effects and $\alpha = -(0.16\pm0.02)z - 0.75\pm0.02$ for the observed sample. These relations are just consistent with each other within the uncertainties, and we find that the inverse Compton scattering from the CMB photons can explain the observed \az\ relation without invoking any environmental or intrinsic differences at high redshift. 

To understand if the final fits were influenced by our choice of magnetic field strength, we repeated the entire experiment with $B=0.1\,$nT and $B=10\,$nT; these are the extreme ends of the expected range of magnetic fields. We found that increasing the magnetic field strength left the slope virtually unchanged, and slightly increased the intercept: $\alpha = -(0.13\pm0.02)z -0.91\pm0.02$. Decreasing the magnetic field strength reduced the slope substantially: $\alpha = -(0.08\pm0.02)z -0.84\pm0.02$. The flattening of the slope for $B=0.1\,$nT is a result of too many sources dropping out of the sample at high redshift as a result of the imposed selection effects. However, this is the extreme end of the magnetic field range and we would only expect a small percentage of sources to have these kinds of magnetic field strengths. Thus for most of the expected range of magnetic field strengths the measured slope of the simulated sample remains the same. 

The scatter of measured spectral indices in the simulated sample is smaller than that in the observed sample, as evident in Fig.~\ref{fig:f4}. The fitted relations agree well above $z\geq1.5$. At lower redshifts the disagreement is due to the fact that there are USS objects at low redshift which come from a positive bias, i.e., searches for USS objects in large, incomplete surveys. The scatter in our simulated sample is more comparable to the scatter in the 3CR survey, which is complete. 

One final selection effect to consider in this analysis is that the sizes of radio galaxies decrease with increasing redshift \citep[e.g.,][]{neeser_linear-size_1995,morabito_investigating_2017}. There is also a dependence of integrated spectral index on size, which \citet{ker_new_2012} have quantified as $\alpha = -0.07$D$-0.94$. Assuming the evolutionary dependence is $D\propto (1+z)^{-n}$ with $n=1.61^{+0.22}_{-0.22}$ \citep{morabito_investigating_2017}, the spectral index would tend to flatten towards higher redshift, but only by $\Delta \alpha \lesssim 0.1$ at $z=5$. This is consistent with the differences we see in the simulated and observed samples. 

By re-simulating a low-redshift radio galaxy population at high redshift, we have effectively assumed there is no intrinsic radio galaxy evolution. It is challenging to decouple observational biases in flux-limited surveys from actual redshift evolution for powerful radio galaxies, although several exemplary studies made progress towards overcoming this challenge \citep[e.g.,][]{neeser_linear-size_1995,blundell_nature_1999,ker_new_2012}. The new LOFAR Two-metre Sky Survey \citep[LoTSS][]{shimwell_lofar_2018} will be crucial for removing these observational biases. The survey will have a 10$\sigma$ limit of 1 mJy and cover the entire Northern sky, providing unprecedented information on both the faint and bright ends of the radio galaxy populations. LoTSS has a resolution well matched to that of the Faint Images of the Radio Sky at 20 cm \citep[FIRST;][]{becker_first_1995}, to aid spectral index studies. In the Southern sky, the combination of the Giant Metrewave Radio Telescope (GMRT) and Square Kilometre Array (SKA) pathfinders like MeerKAT and the Australian SKA Pathfinder (ASKAP) can also provide data to aid spectral index studies. 

\section{Conclusions}
\label{sec:s6}

In this paper we have used spectral modelling of a sample of local bright radio sources to simulate HzRGs and show that the observed \az\ relation can be entirely reproduced by a combination of redshift-dependent inverse Compton losses, coupled with selection effects that are biased towards selecting USS sources from incomplete surveys. 

Finally, we note that while the \az\ relation is still useful for finding candidate high-redshift galaxies, perhaps more high redshift sources could be found by relaxing the strict USS criteria to $\alpha < -0.9$ or $<-0.8$, and coupling the selection with size as suggested by \citet{ker_new_2012}. Ancillary data at near IR wavelengths can be used to exclude low-redshift sources. It is interesting to note that a significant fraction of the radio galaxy population, and potentially some of the most interesting high redshift sources, may fall below even this lower spectral cut. Theoretical models suggest that the contribution from inverse-Compton losses is also a function of source age \citep[e.g.,][]{kaiser_evolutionary_1997,luo_evolution_2010,hardcastle_sim_2018}. Combined with the increased inverse-Compton losses due to redshift, the integrated spectrum of older sources may therefore be dominated by young emission with a diminishing contribution from older (undetectable) lobe plasma. Many large sources at high redshift may therefore have a spectrum more characteristic of particle acceleration regions ($-0.5 > \alpha > -0.7$) and indistinguishable from their younger low redshift counterparts when considering only the integrated radio spectrum. Machine learning techniques that consider multiple source properties to identify high redshift radio galaxies may provide a long term solution for such searches, with the first data release of LoTSS \citep{shimwell_lofar_2018} providing an ideal training set for such methods to be explored.

\section*{Acknowledgements}
We would like to thank the anonymous referee for their very constructive report which helped improve the quality of our manuscript. 
LKM acknowledges financial support from NWO Top LOFAR project, project n. 614.001.006 and the support of the Oxford Hintze Centre for Astrophysical Surveys which is funded through generous support from the Hintze Family Charitable Foundation. This publication arises from research partly funded by the John Fell Oxford University Press (OUP) Research Fund. This research has made use of the NASA/IPAC Extragalactic Database (NED), which is operated by the Jet Propulsion Laboratory, California Institute of Technology, under contract with the National Aeronautics and Space Administration.

\bibliographystyle{mnras}
\bibliography{\myreferences}

\label{lastpage}

\end{document}